\documentclass[a4paper,twocolumn,prb,showpacs,amssymb,aps]{revtex4}
\usepackage{dcolumn}
\usepackage{bm}
\usepackage{graphicx}
\usepackage{xspace}
\def\tc {$T_{\rm C}$}

\newcommand{\sr}{$\mathrm{Sr_{14}Cu_{24}O_{41}}$}
\newcommand{\srca}{$\mathrm{Sr_{14-x}Ca_xCu_{24}O_{41}}$}
\newcommand{\cazwei}{$\mathrm{Sr_{12}Ca_2Cu_{24}O_{41}}$}

\newcommand{\cazwoelf}{$\mathrm{Sr_{2}Ca_{12}Cu_{24}O_{41}}$}
\newcommand{\caelff}{$\mathrm{Sr_{2.5}Ca_{11.5}Cu_{24}O_{41}}$}

\newcommand{\nacu}{$\mathrm{Na_{1+x}CuO_{2}}$}

\newcommand{\mb}{$\mu_B$~}

\newcommand{\figref}[1]{Fig.\,\protect\ref{#1}}

\begin{document}

\title{Magnetisation of hole-doped $\mathbf{CuO_{2}}$ spin chains in
Sr$_{14-x}$Ca$_x$Cu$_{24}$O$_{41}$}
\author{R. Klingeler$^{1,2}$}
\email[]{r.klingeler@ifw-dresden.de}
\author{N. Tristan$^2$}
\author{B. B\"{u}chner$^2$}
\author{M. H\"{u}cker$^{3,4}$}
\author{U. Ammerahl$^{3}$}\author{A. Revcolevschi$^{3}$}

\affiliation{$^1$Laboratoire National des Champs Magn\'{e}tiques Puls\'{e}s, 31432 Toulouse, France.}
\affiliation{$^2$Leibniz-Institute for Solid State and Materials Research IFW Dresden, 01171
Dresden, Germany} \affiliation{$^3$Laboratoire de Physico-Chimie des Solides, Universit\'e
Paris-Sud, 91405 Orsay C\'edex, France}\affiliation{$^4$Physics Department, Brookhaven National
Laboratory, Upton, New York 11973}

\date{\today}

\begin{abstract}

We report on magnetisation measurements of \srca , with $0\leq x \leq 12$, in magnetic fields up to 16\,T.
The low temperature magnetic response of the CuO$_2$ spin chains changes strongly upon doping. For $x=0$, the
ground state with nearly independent dimers is confirmed. Reduction of the number of holes in the chains
through Ca-doping leads to an additional contribution to the magnetisation, which depends linearly on the
magnetic field. Remarkably, the slope of this linear contribution increases with the Ca content. We argue
that antiferromagnetic spin chains do not account for this behaviour but that the hole dynamics might be
involved.

\end{abstract}

 \maketitle

\section{INTRODUCTION}

Introducing charge carriers in a quantum antiferromagnet causes frustration of the magnetic interactions and
degeneracy of the ground state. This degeneracy can be lifted by charge order (CO) which evolves on the
background of the quantum antiferromagnet. One remarkable example is the formation of spatial spin and charge
modulations in the high-T$_C$ cuprates.~\cite{Kivelson03} Another class of low-dimensional cuprates where
spin and charge correlations determine the electronic and magnetic properties are the intrinsically doped
CuO$_2$ spin chain systems which comprise edge-sharing chains like \srca\ and \nacu . In the latter, at
commensurate doping levels, a 1D Wigner lattice occurs where charge order on the chains is determinded by
long-range Coulomb interaction.~\cite{Horsch05} Charge order is also known for the spin chains in \sr . In
the chains of these compounds, there is a non-magnetic dimer ground state related to static charge order on
the chains. The presence of CO is well established e.g. by NMR and ESR data.~\cite{Takigawa98,Kataev01a} Its
presence, however, is intimately connected to the number of holes on the chains. Reducing the number of holes
in the chains by Ca-doping yields a decrease of the stability of CO.~\cite{Kataev01a,Ammerahl00b,Hess04}
Here, we show that the static dimer spin gap $\Delta_d$, which is present in \sr , is even more strongly
affected by the Ca-doping than the CO. We show that in \srca , with $x\neq 0$, at low temperature there is a
finite susceptibility which is not Curie-like. If there is a gap for this additional contribution to the
magnetisation, it is much smaller than the dimer gap $\Delta_d$. Remarkably, susceptibility due to the
low-energy response increases with the Ca content, i.e. the lower the number of holes, the larger the
magnetic susceptibility.

\section{Background}

The compounds \srca\ have an incommensurate layered structure of two alternating
subsystems.~\cite{McCarron88} In these subsystems, two quasi one-dimensional (1D) magnetic structures are
realised which are oriented along the $c$-axis, i.e. $S=\frac{1}{2}$ Cu$_2$O$_3$ spin ladders and CuO$_2$
spin chains. For the low temperature magnetic response, the ladders do not contribute significantly since
they exhibit a non-magnetic ground state and a large spin gap of $\Delta \approx$ 380\,K for the whole doping
series \srca .~\cite{Eccleston98,Katano99,Hess01} Thus, the low temperature magnetic response of these
compounds is determined by the weakly coupled CuO$_2$ spin chains. The chains consist of edge-sharing $\rm
CuO_4$-plaquettes containing Cu$^{2+}$-ions with $S=\frac{1}{2}$ in the undoped case. The Cu-O-Cu bonding
angle amounts to $\sphericalangle \sim 93^{\circ}$.~\cite{McCarron88} For nearest neighbour spins, this
results in a ferromagnetic (FM) superexchange.\cite{Yusha99} The compounds \srca , however, are intrinsically
hole doped with six holes per formula unit. Similar to the high-\tc\ cuprates, the holes are expected to be
mainly of O:2$p$ character and to form Zhang-Rice singlets.~\cite{Zhang88,Nuecker00} In \sr , the holes are
mainly located in the chains.~\cite{Nuecker00} The magnetic coupling of next nearest neighbour Cu-spins via a
hole is antiferromagnetic. Below $T\approx 240$\,K, charge ordering is observed \cite{Takigawa98,Ammerahl00b}
and a dimerized ground state with a spin excitation gap $\Delta \approx$\,130\,K develops.~
\cite{Matsuda99,Regnault99,Ammerahl00b} It is well established that the dimers are formed by second-neighbour
spins separated by a hole. Dimers are separated from each other by two subsequent holes. Another important
aspect of \srca\ is the mutual distortion of the two subsystems due to the incommensurate structure. There is
a pseudoperiodicity of ten chain units for seven ladder units which yields a modulation of each subsystem
with respect to the periodicity of the other one. Since the magnetic interactions are very sensitive to a
variation of the bonding length and the bonding angle, these modulations have a strong impact on the magnetic
properties. It was shown that these modulations govern the low energy properties of the compounds, through
the localization of the magnetic electrons. In particular, the structural distortions favour the formation of
dimers in \sr .~\cite{Gelle04}

Substitution of $\mathrm{Ca^{2+}}$ for $\mathrm{Sr^{2+}}$ leads to a partial charge transfer into the ladder
subsystem owing to chemical pressure.~\cite{Nuecker00,Magishi98} Upon Ca doping, the charge order is
suppressed \cite{Ammerahl00b,Kataev01a,Hess04} and the dimer state becomes unstable
\cite{Regnault99,Matsuda99}. For \srca , with $x \geq 11$, long range antiferromagnetic order is found at low
temperatures ($< 2.5$\,K).~\cite{Isobe99,Akimitsu00} This magnetic order is related to a specific charge
order with alternating spins and holes on the chains.~\cite{Kataev01a} It is a remarkable fact that in this
doping regime the compounds become superconducting upon application of a hydrostatic pressure of
$\sim$3\,GPa.~\cite{Uehara96,Dagotto99}

\section{Experimental and results}

\subsection{Experimental}

We report on magnetisation measurements of \srca\ single crystals, with x=0,2,3,4,5,8,9,11,11.5,12, in
external magnetic fields up to 16\,T. We used a vibrating sample magnetometer (VSM) and a SQUID magnetometer.
We studied crystals of approximately 200\,mg mass grown by the floating zone technique.~\cite{Ammerahl99a}
The samples were cut perpendicular to the $c$-axis (parallel to the chains) and to the $b$-axis
(perpendicular to the CuO$_2$-plaquettes), respectively. Our samples have been previously characterised by
diverse methods, such as magnetisation and thermal expansion \cite{Ammerahl00b}, electrical and thermal
transport \cite{Hess01,Hess04}, inelastic neutron scattering \cite{Regnault99}, and ESR \cite{Kataev01a}.

\subsection{Sr$_{14}$Cu$_{24}$O$_{41}$}

\begin{figure}
\center{\includegraphics [width=1.0\columnwidth,clip] {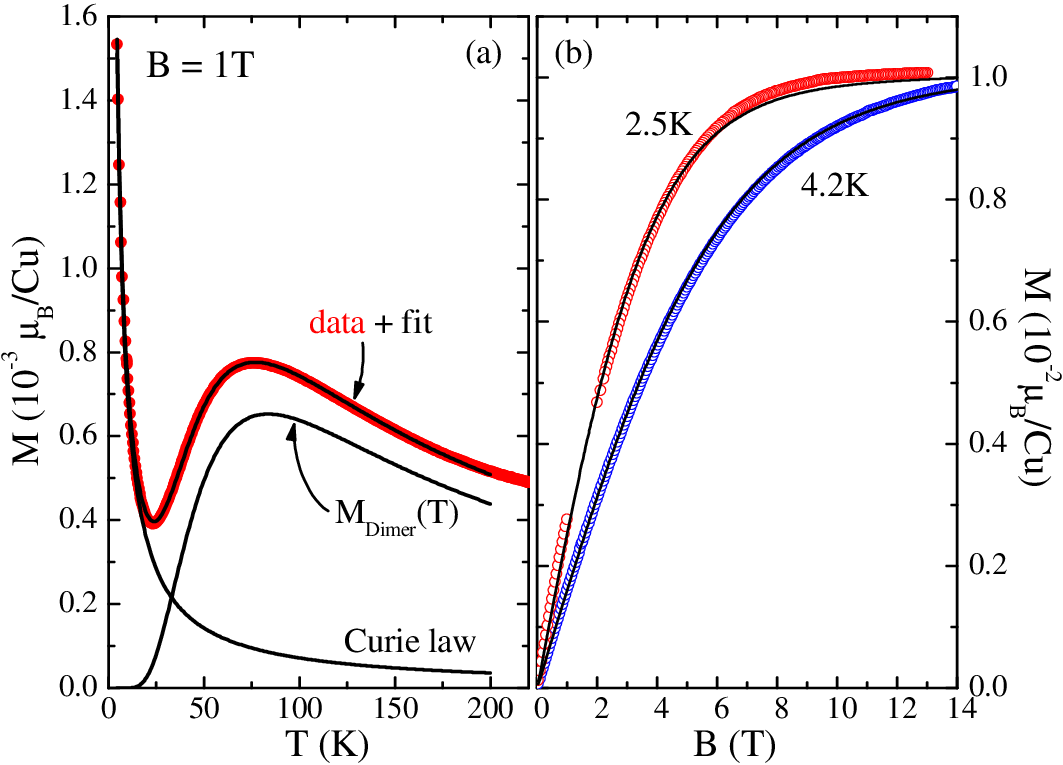}} \caption[]
{\label{sr14}(Colour online) Magnetisation of \sr\ for a constant magnetic field $B=1$\,T versus
temperature (a) and for $T=2.5$\,K ($T=4.2$\,K) versus applied magnetic field (b). $B$ was applied
along the $c$-axis. Lines are fits to the data (see text).}
\end{figure}

In order to recall the magnetic properties of \sr , in Fig.\,\ref{sr14}(a) its magnetisation is shown, in an
applied magnetic field of $B=1$\,T, as a function of temperature. In addition, we show in Fig.\,\ref{sr14}(b)
the magnetisation $M$, at low temperatures, versus $B$. The data confirm the dimerised ground state which has
been observed in our sample by inelastic neutron scattering and thermal expansion
data.~\cite{Regnault99,Ammerahl00b} The data in Fig.\,\ref{sr14}(b) reveal a small value of the magnetisation
of about $1\cdot 10^{-2}$\,\mb /Cu, at $T=2.5$\,K, in a magnetic field of $B=14$\,T. Moreover, the
susceptibility is also very small at high fields, as is evident from the weak dependence of $M$ on $B$ in
this field regime. The data, therefore, show the magnetisation due to about 1\% of free spins while there is
no response from the remaining 99\% of the spins. This agrees with the fact that most of the spins are
dimerised but there are some uncompensated 'defect' spins. Consequently, the data in Fig.\,\ref{sr14}(b) can
be explained in terms of the magnetisation of free spins with $S=\frac{1}{2}$. Applying the Brillouin
function

\begin{equation} M(B) = \chi_{0} \cdot B + \frac{1}{2}N_Sg_c\mu_B \cdot B_{\frac{1}{2}}\left(\frac{g\mu_B(B+\lambda M}{2k_B
T}\right) \label{Brill}, \end{equation}

with the mean field parameter $\lambda$ and the $g$-factor\cite{Kataev01a} $g_c = 2.04$, to the data in
Fig\,\ref{sr14}(b) yields $\chi_{0} \approx 1\cdot 10^{-5}$\,emu/Mol Cu, $\lambda =0$ and $N_S \simeq
0.01$/Cu. Here, the linear term $\chi_0\cdot B$ considers the Van-Vleck magnetism of the Cu-ions. The data in
Fig.\,\ref{sr14}(b) provide the number $N_S$ of free spins in our sample very accurately, which allows to
estimate their response versus temperature. Considering the Curie contribution due to the free spins and
applying the model of independent dimers (cf. Refs.\,[\onlinecite{Carter96,Ammerahl00b}]) to the data in
Fig.\ref{sr14}(a) allows to extract the dimer spin gap and the number $N_D$ of dimers. The analysis yields a
gap of $\Delta_d = 134$\,K and $N_D = 0.0738$/Cu. In this analysis, the number of magnetic Cu-sites amounts
to $N_S + 2N_D \approx 3.78$/f.u. in our \sr\ sample.

\subsection{Sr$_{14-x}$Ca$_x$Cu$_{24}$O$_{41}$ ($0\leq x \leq 12$)}

It is known that application of chemical pressure through Ca-doping leads to a hole transfer from the chains
to the ladders in \srca , thereby destabilizing the charge ordered ground state which is present for
$x=0$.~\cite{Nuecker00,Ammerahl00b,Kataev01a} In our study, we concentrate on the effect of Ca-doping on the
low temperature magnetic susceptibility. We note, that the magnetic properties have been studied previously
by investigating the temperature dependence of the magnetisation of polycrystals at $B=1$\,T.~\cite{Carter96}
These data of \srca , with $0\leq x \leq 10$, have been analysed in terms of the free dimer model.
Qualitatively, our $M(T)$ data on single crystals are similar to those in e.g.
Refs.\,[\onlinecite{Carter96,Kumagai00}].

\begin{figure}
\center{\includegraphics [width=1.0\columnwidth,clip] {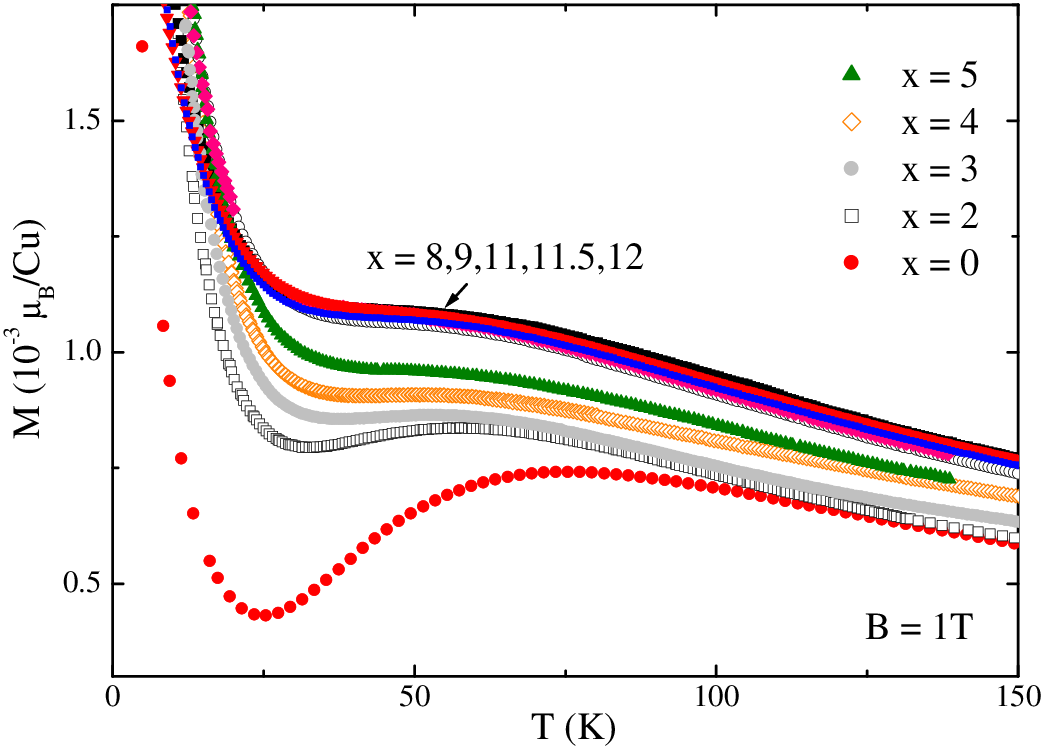}} \caption[]
{\label{mvt}(Colour online) Magnetisation of \srca\ with $0\leq x\leq 12$ in a magnetic field of
$B=1$\,T parallel to the $c$-axis.}
\end{figure}

Fig.\,\ref{mvt} displays the magnetisation data at small magnetic field for \srca\ with $0 \leq x \leq 12$.
Obviously, changing the Ca-content significantly affects the properties at low temperatures. In particular,
the signature of the dimer gap, a broad maximum of $M$ around 75\,K, which is very pronounced in \sr ,
becomes much weaker upon Ca doping.  The data show that the magnetisation of \srca , at $T\gtrsim 25$\,K,
increases upon doping for $x\leq 8$ and is nearly independent of the Ca content for larger $x$.  Due to the
fact that upon Ca doping, however, the Curie contribution significantly increases, these data do not answer
the question of the existence of a spin gap for $x>0$. In particular, it is possible to describe perfectly
the data for $0\leq x\leq 5$ in terms of the free dimer model by including the Van-Vleck magnetism and the
Curie contribution due to free spins. However, our magnetisation data $M$ vs. $B$, which we present in the
following, show that the dimer model is not valid for $x\neq 0$.

\begin{figure}
\center{\includegraphics [width=1.0\columnwidth,clip] {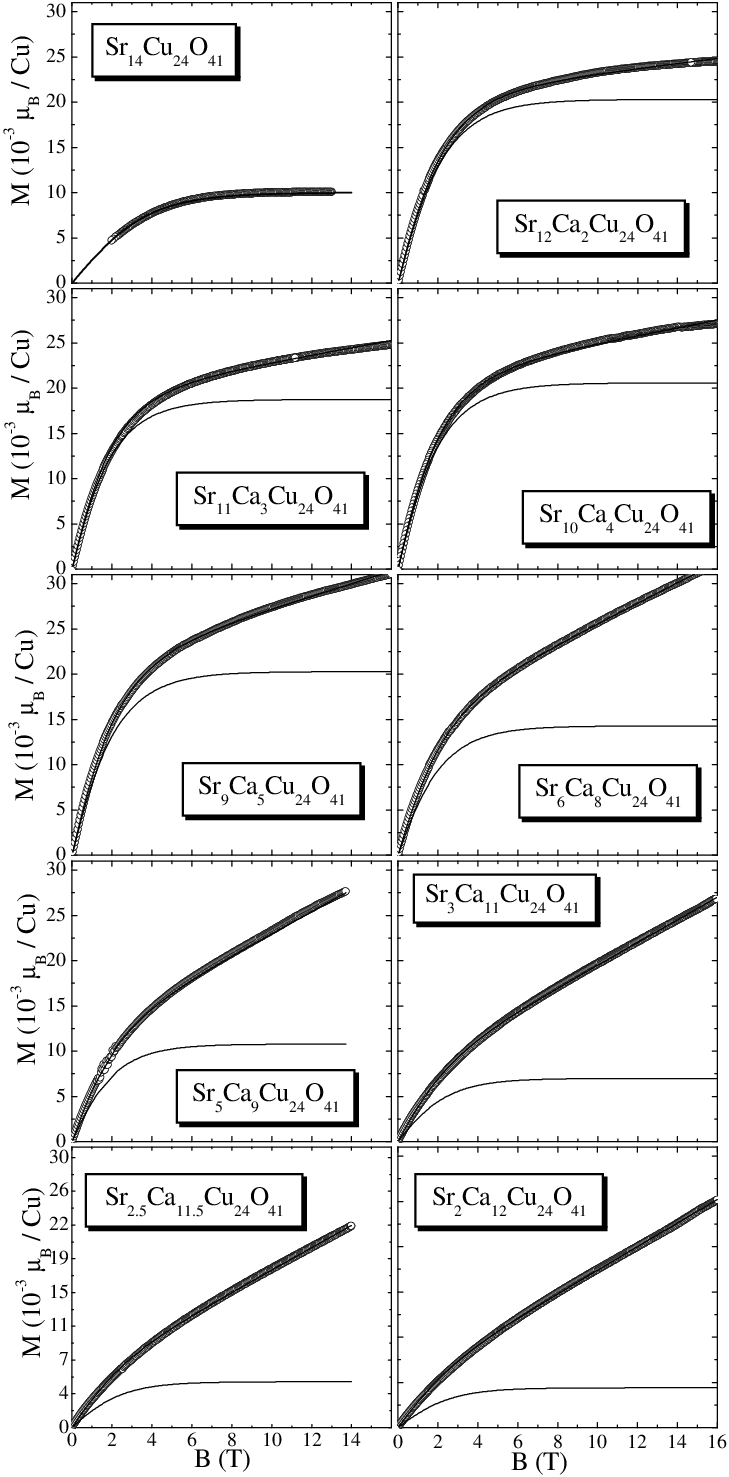}} \caption[]
{\label{mvb}Magnetic field dependence of the magnetisation of \srca\ with $0\leq x \leq 12$ at
$T=2.5$\,K. The lines describe the response of nearly free spins (see text).}
\end{figure}

\begin{figure}
\center{\includegraphics [width=1.0\columnwidth,clip] {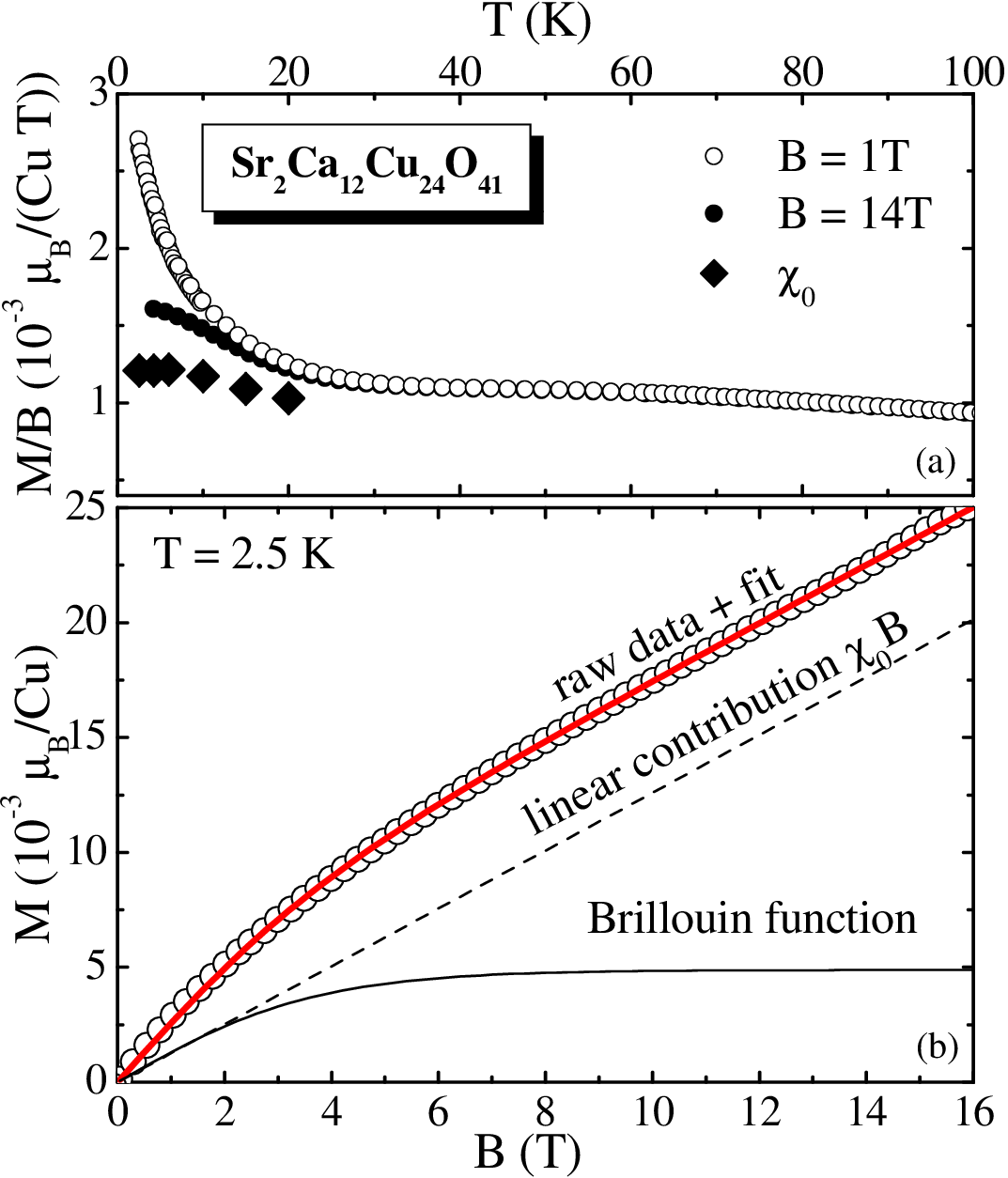}} \caption[] {\label{ca11}(Colour
online) Magnetisation of \cazwoelf . (a) Temperature dependence of $M/B$ for $B=1$\,T and $B=14$\,T. (b)
Field dependence of $M$ at $T=2.5$\,K (cf. Fig.\,\ref{mvb}). The data are well described by a fitting
function which considers a linear term (dashed line) and a Brillouin function (straight line). The diamonds
in (a) correspond to the susceptibility $\partial M/\partial B$ of the linear term in (b).}
\end{figure}

In order to clarify the origin of the low temperature magnetisation in \srca , we studied the field
dependence of $M$ at $T=2.5$\,K. The data are displayed in Fig.\,\ref{mvb}. As already shown in
Fig.\,\ref{sr14}, for $x=0$ the magnetisation in high magnetic fields is nearly independent of $B$, i.e. the
susceptibility $\partial M/\partial B$ is almost zero, indicating the spin gapped state. When the number of
holes in the chains, however, is reduced through the Ca-doping, the data show a finite slope of $M(B)$ in
high magnetic fields. In particular, a finite slope is already observed for $x=2$. Remarkably, the slope of
$M(B)$ increases monotonically upon Ca-doping. This becomes even more apparent if the data are analysed in
terms of Eq.\,\ref{Brill}. The fitting procedure is illustrated in Fig.\,\ref{ca11}(b) for the example of
\cazwoelf . The experimental data are described in terms of a Brillouin function and a linear term as
displayed in the figure by the solid and the dashed line, respectively. The fitting yields a reliable result
in the whole field range. We mention that, in contrast with \sr , for the sample with $x=12$, the Brillouin
function does not describe entirely independent spins but a small mean field parameter of $\lambda \simeq
60$\,(emu/Mol Cu)$^{-1}$ is necessary to describe the data correctly. In order to demonstrate the effect of
the magnetic field, also the temperature dependence of the static susceptibility $M/B$ of \cazwoelf\ for
$B=1$\,T and $B=14$\,T is shown in Fig.\,\ref{ca11}(a). These data confirm that, below $\sim$20\,K, the field
dependence of $M$ is non-linear. For comparison, the susceptibility $\chi_0$ which corresponds to the linear
term in Eq.\,\ref{Brill} is plotted in Fig.\,\ref{ca11}(a). We emphasize that, in the case $x=12$,
$\chi_0=\partial M_{lin}/\partial B$ is much larger than for \sr . Thus, in contrast with the undoped
compound, where $\chi_0$ is ascribed completely to the Van-Vleck magnetism \footnote{$\chi_{VV} \approx 1.2
\cdot 10^{-5}$\,emu/Mol Cu for $B||c$.}, the linear term in Eq.\,\ref{Brill} has a different physical origin.
We also extracted $\chi_0$ from $M$ vs. $B$ measurements at higher temperatures as it is illustrated by the
diamonds in Fig.\,\ref{ca11}(a). The data imply that $\chi_0$ is only very weakly temperature dependent for
$T\leq 20$\,K.~\footnote{To subtract the susceptibility of the Curie-like spins, we applied Eq.\,\ref{Brill}
and considered $N_S\simeq 5\cdot 10^{-3}$ spins/Cu to be temperature independent. In order to minimize the
number of fitting parameters, we set $\lambda =0$ since this does not affect the resulting value $M_{lin}/B$
significantly.}

\begin{figure}
\center{\includegraphics [width=1.0\columnwidth,clip] {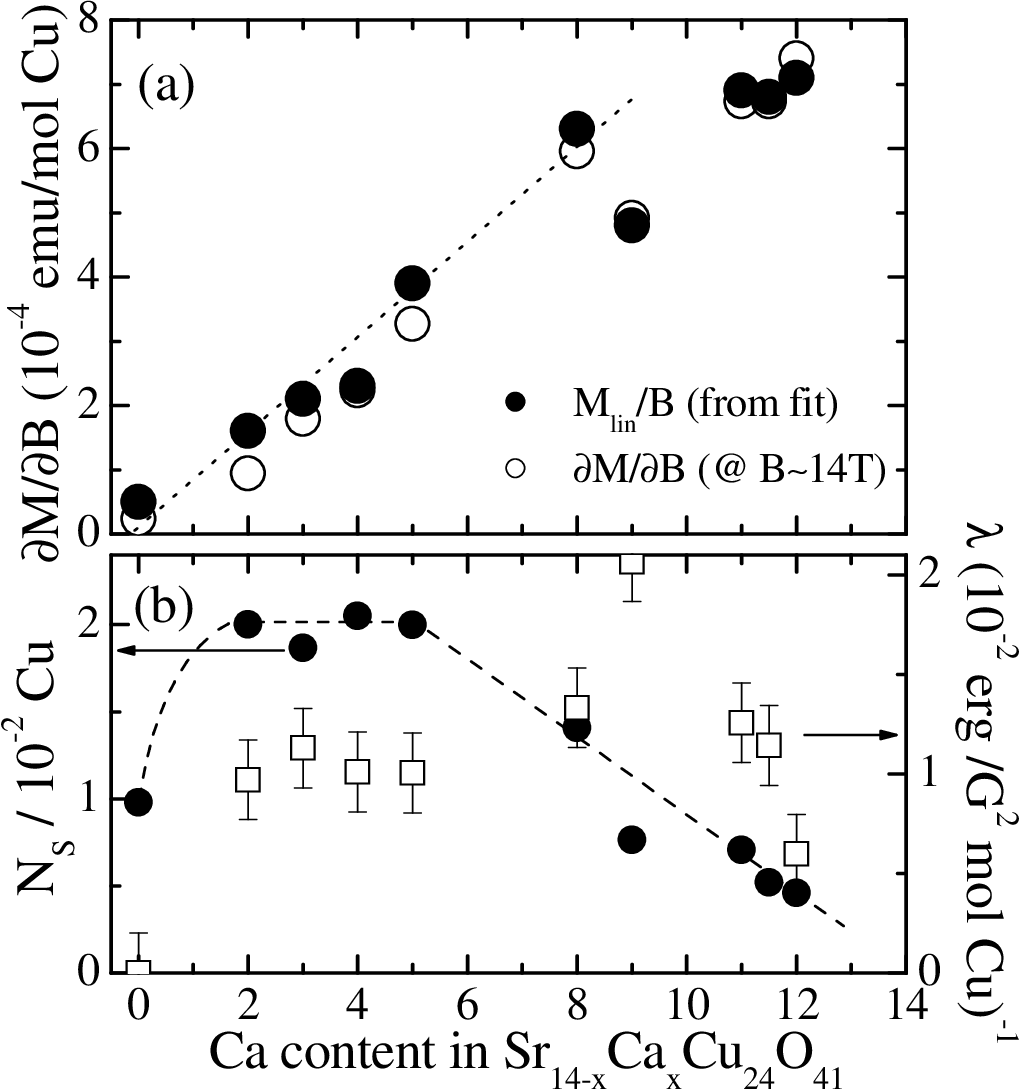}} \caption[] {\label{fit}Results of
fitting the data of Fig.\,\ref{mvb} with Eq.\,\ref{Brill}. (a) Linear term $\chi_0 = M_{lin}/B$ from fit
(full circles) and derivative of the raw data $\partial M/\partial B$ at $B\sim 14$\,T (open circles). (b)
Number of quasi-free spins $N_S$ (circles) and mean-field parameter $\lambda$ (squares). The lines are guides
to the eye.}
\end{figure}

The above-described fitting procedure has been applied to the $M(B)$ curves of all compositions under study.
The resulting Brillouin function is displayed in Fig.\,\ref{mvb} for each composition. It is apparent that
the difference between the data and the Brillouin function, i.e. the linear contribution to $M(B)$, increases
monotonically upon Ca-doping. This major result of our study is visualized in Fig.\,\ref{fit}(a), where the
susceptibility $\chi_0$ from Eq.\,\ref{Brill} is plotted. In addition, we show the field derivative of the
experimental magnetisation data $\partial M/\partial B$ at $B \sim 14$\,T. For $x\neq 0$, obviously, $\chi_0$
is much larger than the Van-Vleck magnetism. Hence, both the raw data and our analysis clearly show that, for
$x \neq 0$, there is an additional magnetic contribution which gives rise to a linear field dependence of the
magnetisation in the accessible field range. This response is due neither to free spins nor to dimers. The
free spins are nearly completely aligned in high magnetic fields, at low temperature. The spin gap of the
dimers is much larger than the applied magnetic field, i.e. $\Delta_d /(g\mu_{\rm B}) \gg 14$\,T.

\begin{figure}
\center{\includegraphics [width=1.0\columnwidth,clip] {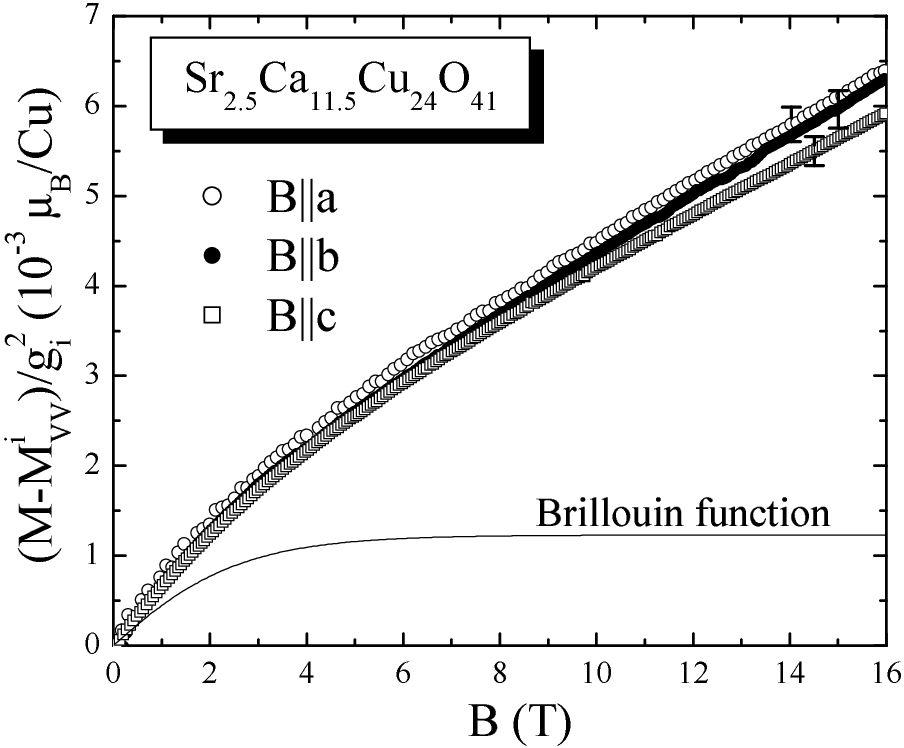}} \caption[]
{\label{ani115}Magnetisation of \caelff\ for $B$ parallel to the $a$, $b$, and $c$-axis, respectively. Data
are corrected by the anisotropy of the $g$-factor and the Van-Vleck magnetism. Error bars show the
uncertainty due to the correction.}
\end{figure}

The unusually large value of $\chi_0$ is almost isotropic. This is illustrated in \figref{ani115}, where the
magnetisation of \caelff , at 2.5\,K, is shown for $B$ parallel to the $a$, $b$, and $c$-axis, respectively.
In order to study the anisotropy of $\chi_0$, the data are corrected for the anisotropy of the $g$-factor and
of the Van-Vleck magnetism. Values of the $g$-factor were taken from Ref.\,[\onlinecite{Kataev01a}].
Obviously, the magnetisation is very similar for all field directions. The data imply $\Delta M/g_i^2 <
0.01$\mb /Cu, at 16\,T, which is comparable to the error which originates from the uncertainty of the
$g$-factor.

In addition to the surprising strong dependence of $\chi_0$ on the Ca content, our analysis yields the doping
dependence of the number of the quasi-free spins. These data are shown in Fig.\,\ref{fit}(b). Starting from
$x=0$, $N_S$ increases sharply for $x=2$ and stays constant for $2\lesssim x\lesssim 5$. For higher
Ca-content, the amount of quasi-free spins decreases. This result confirms a study on polycrystalline samples
with $0\leq x \leq 6$ and $x\geq 12.6$, where a similar doping dependence but larger values of $N_S$ were
observed.~\cite{Isobe99} As already mentioned, our analysis suggests that the spins which give rise to the
Curie-like contribution are not completely free for $x\neq 0$ but they are ferromagnetically coupled. As
displayed in Fig.\,\ref{fit}(b), the mean-field parameter which we extract from the data is of the order of
$\lambda \simeq 100$\,(emu/Mol Cu)$^{-1}$ for $x\neq 0$.

The central result of our experimental study is summarized in Fig.\,\ref{fit}(a). For \sr , our data confirm
the presence of the spin gap. For $x\neq 0$, however, we find a finite susceptibility $\chi_0$ at $T=2.5$\,K
which is not due to free spins. This unusual magnetic response is practically isotropic and increases
linearly upon Ca-doping in the range $0\leq x \leq 9$. For the example of $x=12$, it is only weakly
temperature dependent.

We mention that our data do not prove that $\chi_0$ is gapless. Our measurements at $T=2.5$\,K would not
detect a very small gap. Moreover, despite the fact that Eq.\,\ref{Brill} provides a reliable description of
the data in Fig.\,\ref{ca11}, in addition to the strong field dependence of the Brillouin function there
might be a weak field dependence of $\chi_0$ at low magnetic fields, too. It is obvious, however, that at
high fields the linear term in Eq.\,\ref{Brill} describes the data correctly. Therefore, if a gap $\Delta_0$
for $\chi_0$ does exist, we estimate from our data that $\Delta_0$ is of the order of only several Kelvin.

\section{DISCUSSION}

\begin{figure}
\center{\includegraphics [width=1.0\columnwidth,clip] {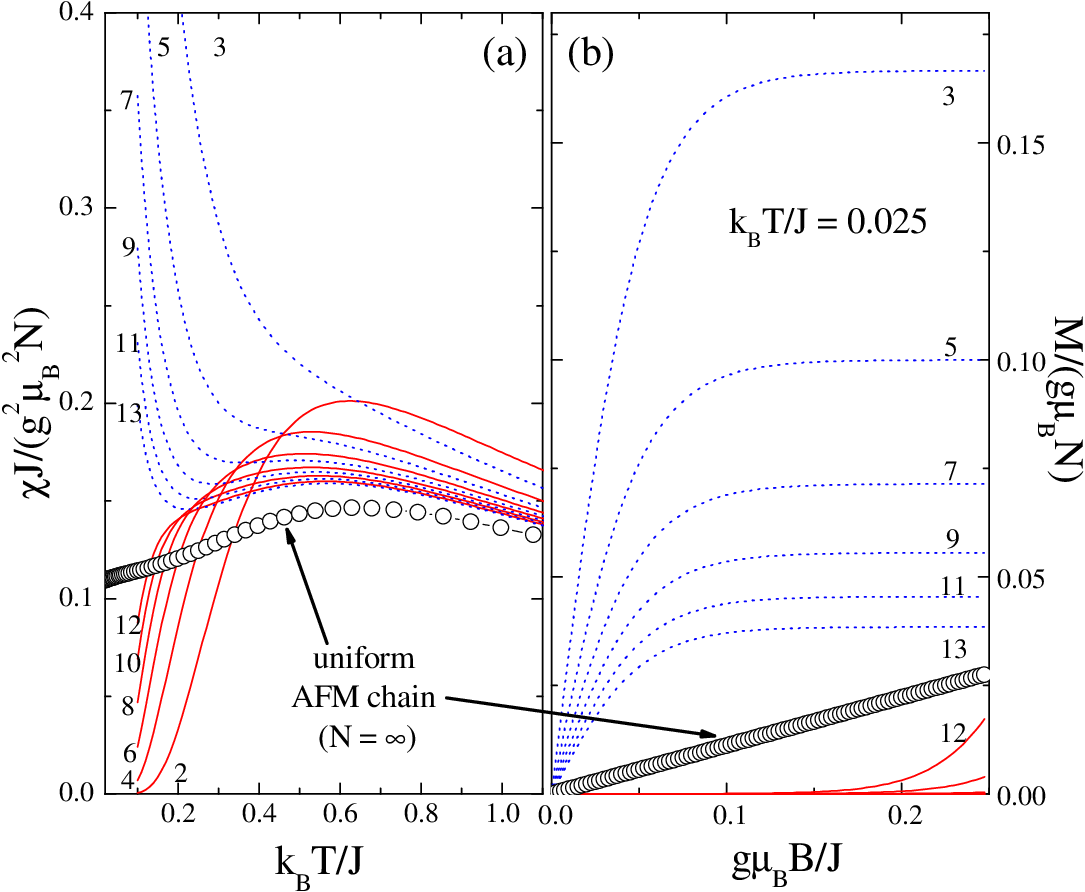}} \caption[] {\label{finite}(Colour
online) Susceptibility of finite $S=1/2$ spin chains with $2\leq N \leq 13$ sites (a) versus temperature and
(b) at $k_{\rm B}T/J = 0.025$ versus external magnetic field.~\cite{Honecker} For comparison, the data for a
uniform AFM spin chain are plotted.~\cite{Kluemper}}
\end{figure}

The origin of the finite low-energy magnetic contribution $\chi_0$ is not obvious. Since it is known that
Ca-doping reduces the number of holes in the chains, it is straightforward to assume that finite
antiferromagnetic spin chains evolve upon Ca-doping. The occurrence of finite spin chains, however, can not
explain our data. This is shown by exact diagonalisation studies which calculate the magnetic response of
antiferromagnetic $S=1/2$ spin chains with $2\leq N \leq 13$ sites.~\cite{Honecker} The temperature
dependence of $\chi$, which is displayed in Fig.\,\ref{finite}, shows that for finite spin chains there is
still a spin excitation gap. For chains with an even number of sites, the susceptibility $\partial M/\partial
B$, at low temperature, becomes zero. For odd chains, there is a Curie-like upturn due to the uncompensated
spin. Fig.\,\ref{finite}(a) once more illustrates that, in the presence of inevitable crystal defects which
also yield to a Curie-like upturn even in the case of the uniform antiferromagnetic chain, it is not possible
to detect finite susceptibility at low temperatures experimentally by $M(T)$ measurements in small magnetic
fields. On the other hand, for even as well as for odd spin chains, at low temperatures, the susceptibility
becomes essentially zero for larger magnetic fields $0 \ll g \mu_{\rm B}B/J \ll 1$, as can be seen in
Fig.\,\ref{finite}(b). The data show that, at $g \mu_{\rm B}B/J \simeq 0.15 \simeq 15$\,T, the magnetic field
is large enough to align the residual spins in the odd chains but too small to overcome the spin gap. Hence,
finite spin chains do not account for our experimental observation of $\chi_0 \neq 0$ in Fig.\,\ref{fit}(a),
which evidences a finite slope in $M(B)$ at high magnetic fields. We mention that recent numerical results on
the electronic structure of \srca\ in Refs.\,[\onlinecite{Gelle04,Gelle05}] also do not explain the observed
$\chi_0$.

For higher Ca-content, $x\sim 12$, the scenario of an antiferromagnetic Heisenberg chain was
suggested.~\cite{Kataev01a} This scenario is based on the fact that for $x\sim 12$ one might expect a
reduction of the hole content to $\sim$50\% in the chains, due to the charge transfer into the ladders. In
this case, it was suggested that an antiferromagnetic chain of next nearest neighbour spins develops. The
presence of a uniform antiferromagnetic spin chain straightforwardly explains the observation of $\chi_0\neq
0$ since the spin excitations of the Heisenberg chain are gapless (cf. Fig.\,\ref{finite}). Our data,
however, cannot prove this scenario because we cannot describe both the temperature dependence and the field
dependence of the magnetisation in terms of a uniform Heisenberg chain by utilizing a common set of
parameters for any compound under study. Moreover, the scenario of the Heisenberg chain should not account
for $\chi_0\neq 0$ at very small Ca-contents. We recall that we observe an additional magnetic contribution
$\chi_0$ even for \cazwei , where the number of holes is not supposed to be much smaller than in \sr . For
$x=2$, a uniform Heisenberg chain is very unlikely and, in this case, another mechanism must account for a
finite $\chi_0$. We hence conclude that a simple 1D spin model based on either finite AFM spin chains or the
uniform AFM Heisenberg chain does not explain the observed doping dependence of $\chi_0$.

We now briefly discuss the fitting parameters $N_S$ and $\lambda$ in Eq. \ref{Brill} [cf. Fig. \ref{fit}(b)].
According to Ref.\,[\onlinecite{Gelle04}], for $x=0$, the Curie-like response is due to $N_S$ single spins
which are very weakly coupled to neighbored dimers. Introducing Ca-ions leads to a smaller amount of holes in
the chains. Thus, for small a Ca-content, $N_S$ (i.e. the number of single spins) increases. This trend,
however, reverses for $x>5$. Despite the fact that charge order and dimers are progressively destabilized
\cite{Kataev01a,Ammerahl00b,Hess04}, the Curie-like response decreases for $x>5$ ($N_S$ decreases). As
visible from numerical results in Fig.\,\ref{finite}(b), however, from our magnetisation data we cannot
distinguish whether the Curie-like response is due to single free spins ('monomers') or due to the same
number of trimers, quintumers, etc. For example, one expects similar magnetic response for monomers and for a
three times larger number of spins which are coupled to trimers. The parameter $N_S$ hence measures the sum
of all finite chain fragments with a Curie-like response, i.e. of finite chains with an odd number of spins.
The fact that $N_S$ decreases for $x>5$ therefore implies that defects (i.e. defects of the dimerized state)
are not only individual spins but tend to form antiferromagnetic chain fragments. This assumption is
confirmed by a recent numerical study\cite{Gelle05} which suggests the absence of individual free spins for
\srca\ with $x= 13.6$ but implies larger spin cluster. In particular, the low temperature spin configuration
shown in Ref.\,[\onlinecite{Gelle05}] comprises neither single free spins nor spin clusters with an odd
number of spins for $x=13.6$. This does not contradict our data since we extrapolate $N_S\sim 0.1$, for this
doping level, which might be attributed to inevitable crystal imperfections. We mention, that the decrease of
$N_S$, for larger $x$, strongly confirms that, for $x=0$, the Curie-like response is not due to crystal
imperfections but due to defects of the dimer state since the former should increase continuously upon
Ca-doping. Interestingly, the defects are ferromagnetically correlated in the case of $x\neq 0$, i.e. the
mean-field parameter in Eq.\,\ref{Brill} $\lambda > 0$ [see Fig.\,\ref{fit}(b)]. The observation $\lambda\neq
0$ means that the defects of the dimer state, which at least for $x>5$ contain finite chain fragments with an
odd number of spins (e.g. trimers), magnetically interact with each other. Quantitatively, applying

\begin{equation} \lambda =
\frac{2\tilde{J}}{N_Ag^2\mu_B^2}\end{equation}

yields a magnetic coupling of the order of $k_B\tilde{J} \sim 80$\,K. Since the next nearest neighbour
coupling via one hole is antiferromagnetic and the coupling via two holes is only weakly ferromagnetic
\cite{Regnault99}, the data thus suggest that the nearest neighbour coupling accounts for the observed
ferromagnetic correlations. We therefore suggest that, for $x\neq 0$, finite chains with an odd number of
spins reside on nearest neighbour chain fragments.

While the fitting parameters $N_S$ and $\lambda$ in Eq. \ref{Brill} can be explained straightforwardly, we
are not aware of any 1D spin model which can explain the unusual doping dependence of $\chi_0$. It might be
necessary to apply more advanced spin models, which, e.g., include the interchain coupling. We recall,
however, that the additional contribution to the magnetisation is isotropic. One might also speculate whether
charge degrees of freedom are involved. This assumption is based on the observed doping dependence of
$\chi_0$. Fig.\ref{fit}(a) suggests a linear dependence of $\chi_0$ on the Ca-content up to $x \approx 8$.
For larger $x$, $\chi_0$ saturates. We recall that the dimerised ground state in \sr\ is associated with a
particular charge order. Ca-doping leads to a charge transfer into the ladder subsystem, thereby
destabilizing the charge order. It has been shown, by ESR \cite{Kataev01a}, transport \cite{Hess04} and
thermal expansion data \cite{Ammerahl00b}, that the charge ordering becomes progressively suppressed with
increasing $x$ and, for $x>8$, the CO seems to disappear. Apparently, the increase of $\chi_0$ is associated
with the destabilization of the dimer/charge ordered state upon Ca-doping. This implies that the finite
susceptibility $\chi_0$ at 2.5\,K cannot coexist with the charge ordering and charge mobility might be
crucial for the observed phenomenon. On the other hand, our observation of $\chi_0 \neq 0$, at $T=2.5$\,K,
hence suggests a significant perturbation of the charge ordering, even at lowest temperatures.

\section{Conclusion}

The analysis of the temperature dependence of the magnetisation of \srca\ in $B=1$\,T alone does not provide
a correct understanding of the low temperature magnetisation. Whereas, the high magnetic field dependence of
$M$ at low temperatures is crucial. Our study of the high field magnetisation confirms the spin gap in the
CuO$_2$ spin chains of \sr . The low temperature magnetic response of the CuO$_2$ spin chains, however,
strongly changes when the number of holes in the chains is reduced through Ca-doping. We find an additional
isotropic contribution to the magnetisation, which linearly depends on magnetic fields between $\sim 3$\,T
and 16 T. Our data give evidence that, in \srca , with $x\neq 0$, at low temperature there is a finite
susceptibility which is due neither to free spins nor to dimers. If there is a gap for this response, it is
much smaller than the dimer gap $\Delta_d$. Remarkably, the slope of the linear contribution increases with
the Ca content, i.e. with the reduction of the number of holes in the chains. We argue that antiferromagnetic
spin chains do not account for this behaviour and possible field induced dynamics of the holes might be
relevant.

\begin{acknowledgments}
We thank A. Honecker for calculating the response of finite spin chains and A. Kl\"{u}mper for providing the
data for the uniform AFM Heisenberg chain. We thank V. Kataev, C. Hess and T. Woodcock for critically reading
the manuscript and for fruitful discussions. This work was supported by the Deutsche Forschungsgemeinschaft
(DFG) within SPP 1073 (BU 887/1-3). R.K. acknowledges support by the DFG through KL 1824/1-1.
\end{acknowledgments}

\end{document}